\documentstyle[preprint,aps]{revtex}
\begin{document}
\preprint{UCM-37-1994}
\title{Green function approach to interface states in band-inverted
junctions}

\author{Francisco Dom\'{\i}nguez-Adame}

\address{Departamento de F\'{\i}sica de Materiales,
Facultad de F\'{\i}sicas, Universidad Complutense,
E-28040 Madrid, Spain}

\maketitle


\narrowtext

Narrow-gap semiconductor compounds like Pb$_{1-x}$Sn$_x$Te and
Hg$_{1-x}$Cd$_x$Te present band inversion under compositional variation.
In a band-inverted heterojunction the fundamental gap, defined as the
difference between $\Gamma_6$ and $\Gamma_8$ energies, has opposite
signs on each side \cite{Seeger}.  Type III superlattices with band
inversion of CdTe/HgTe and PbTe/SnTe has been successfully grown in the
past \cite{Faurie,Ishida}.  One of the most conspicuous characteristic
of band-inverted heterojunctions is the existence of interface states
lying within the fundamental gap, provided that the two gaps overlap
\cite{Pan1,Korenman,Agassi,Pan2,Litvinov}.  In IV-VI compounds those
interface states are properly decribed by means of a two-band model
using the effective ${\bf k}\cdot {\bf p}$ approximation.  On the
contrary, the analysis is more complex in II-VI compounds due to mixing
with heavy-hole states since non centro-symmetry effects are not
negligible in this case.  The equation governing conduction- and
valence-band envelope-functions in a simple two-band model, neglecting
far-band corrections, is a Dirac-like equation.  Exact solutions can
then be found in view of this analogy because one can use elaborated
techniques like those related to supersymmetric quantum mechanics
\cite{Pan2}.  The aim of this paper is to present an alternative way of
solution based on the so called {\em point interaction potentials}
\cite{JPA1,JPA2} (any arbitrary sharply peaked potential approaching the
$\delta$-function limit) along with a Green function method.  We believe
that our treatment gives a very intuitive explanation of the origin of
interfaces states while other approaches may obscure the way how those
states arise.  Moreover, the effects of external electric and magnetic
fields can be included in a straightforward fashion, as we will show
later.

In the effective-mass approximation the electronic wave function is a
sum of products of Bloch functions at the band-edge with slowly varying
envelope-functions. The two-band model Hamiltonian in the absence of
external fields is of the form
\begin{equation}
{\cal H}=v_{\perp}\alpha_yp_y+v_z\alpha_zp_z+{1\over 2}\,\beta E_G(z),
\label{1}
\end{equation}
where the $Z$ axis is perpendicular to the heterojunction, $E_G(z)$
stands for the position dependent gap, $\alpha_y$, $\alpha_z$ and
$\beta$ are the usual $4\times 4$ Dirac matrices, $v_\perp$ and $v_z$
are interband matrix elements having dimensions of velocity.  As usual,
it is assumed that these matrix elements are constant through the whole
heterostructure due to the similarity of the zone centre in both
semiconductors.  Since the gap depends only upon $z$, the transversal
momentum is a constant of motion and we can set the $Y$ axis parallel to
this component.  In the two-band case there are four envelope-functions
including spin and we arrange them in a four component vector $F({\bf
r})$.  This vector satisfies the equation
\begin{equation}
{\cal H}\,F({\bf r}) = [E-V(z)]\,F({\bf r}),
\label{2}
\end{equation}
where $V(z)$ gives the position of the gap centre.  It is understood
that the growth direction is $[111]$.  The way $V(z)$ changes from one
layer to another is not well understood but, assuming that the interface
states spread over distances much larger than the interface region, we
can confidently consider it as a step-like function.  Accordingly we
take
\begin{mathletters}
\label{3}
\begin{eqnarray}
E_G(z)&=&E_{GL}\theta(-z)+E_{GR}\theta(z), \\
V(z)&=&V_{L}\theta(-z)+V_{R}\theta(z),
\end{eqnarray}
\end{mathletters}
$\theta$ being the Heaviside step function.  Here, the subscripts $L$
and $R$ mean left and right sides of the heterojunction, respectively.

As we have already mentioned above, the momentum perpendicular to the
interface is conserved, and therefore we look for solutions of the form
\begin{equation}
F({\bf r})=F(z)\exp \left( {i\over \hbar}\, {\bf r}_\perp\cdot{\bf
p}_\perp \right)
\label{4}
\end{equation}
to Eq.~(\ref{2}). The function $F(z)$ satisfies the following equation
\begin{equation}
\left(\alpha_y v_\perp p_\perp + \alpha_z v_z p_z+{1\over 2}\,\beta
E_G(z)-E+V(z)\right)\,F(z)=0.
\label{5}
\end{equation}
A simple way to solve this equation is the Feynman-Gell-Mann {\em ansatz}
\cite{Feynman}
\begin{equation}
F(z)=\left(\alpha_y v_\perp p_\perp + \alpha_z v_z p_z+{1\over 2}\,\beta
E_G(z)+E-V(z)\right)\,\chi(z).
\label{6}
\end{equation}
After a little algebra we obtain
\begin{equation}
\left\{ -\,{d^2\phantom{z^2}\over
dz^2}+{1\over \hbar^2v_z^2} \left[ {1\over 4}
\,E_G(z)^2-[E-V(z)]^2+v_\perp^2p_\perp^2\right]-i\Delta
\alpha_z(\beta-\lambda)\delta(z)\right\} \chi(z)=0.
\label{7}
\end{equation}
For brevity we have defined
\begin{mathletters}
\label{8}
\begin{eqnarray}
\Delta&=&{E_{GR}-E_{GL}\over 2\hbar v_z}, \\
\lambda&=&2\,{V_R-V_L\over E_{GR}-E_{GL}},
\end{eqnarray}
\end{mathletters}
and we have used the relationships $d\theta(\pm z)/dz=\pm \delta(z)$.
Note that in the case of band-inverted heterojunction $E_{GR}E_{GL}<0$.

It is worth mentioning that Eq.~(\ref{7}) is nothing but a Klein-Gordon
equation with scalar-like and electrostatic-like terms depending on
position (like a relativistic spinless particle with a
position-dependent mass in an electric field as occurs in QED) plus a
point interaction potential arising from the discontinuity of the gap
and the gap centre.  The occurrence of this short-range potential makes
it possible the existence of bound states deep in the gap.  In order to
find the bound states we use a Green function formalism, similar to
previously used in the case of the Dirac equation with point interaction
potentials \cite{JPA3}.  To this end, let us consider the Green function
associated to Eq.~(\ref{7}) without the point interaction potential
\begin{equation}
\left\{ -\,{\partial^2\phantom{z^2}\over
\partial z^2}+{1\over \hbar^2v_z^2} \left[ {1\over 4}
\,E_G(z)^2-[E-V(z)]^2+v_\perp^2p_\perp^2\right]
\right\} G(z,z';E)=I_4 \delta(z-z'),
\label{9}
\end{equation}
where $I_4$ stands for the $4\times 4$ unity matrix. Since we are
interested in bound state levels, boundary conditions reads
\begin{equation}
\lim_{|z|,|z'|\to\infty}G(z,z;E)=0.
\label{10}
\end{equation}
The Green function is a $4\times 4$ matrix which permits the
factorization $G(z,z';E)=g(z,z';E)I_4$, where $g(z,z';E)$ is a scalar
function since the operator in the left hand side of Eq.~(\ref{9})
is scalar.

The solution of Eq.~(\ref{7}) is then simply written out as follows
\begin{eqnarray}
\chi(z)&=&i\Delta\alpha_z(\beta-\lambda)\mbox{$\int_{-\infty}^\infty$}
\>dz'\,G(z,z';E) \delta(z') \chi(z'),\nonumber\\
&=&i\Delta\alpha_z(\beta-\lambda)g(z,0;E)\chi(0).
\label{11}
\end{eqnarray}
It is assumed that the envelope-functions are continuous at the
heterojunction so that the value $\chi(0)$ is defined without
ambiguities;  this is completely different to what found in the Dirac
equation for point intereaction potentials (see Ref.~\cite{JPA1} and
references therein).  Once the Green function is known, the $4$-vector
$\chi(z)$ can be obtain and using Eqs.~(\ref{4}) and (\ref{6}) the
envelope-functions are finally determined.  Bound state levels can be
computed taking the limit $z\to 0$ in Eq.~(\ref{11}).  To obtain
nontrivial solutions we require the $4\times 4$ determinant to vanish.
Thus, using the definitions of $\Delta$ and $\lambda$ given in (\ref{8})
we obtain
\begin{equation}
{1\over \hbar^2v_z^2}\,\left[{1\over 4}\,(E_{GR}-E_{GL})^2-
(V_R-V_L)^2 \right]=\left( {1\over g(0,0;E)}\right)^2.
\label{12}
\end{equation}
At this point we would like to stress that we just require the value of
the Green function at the origin of the $(X,X')$ plane if only the bound
state levels are needed.  In the absence of external fields, as we are
considering here, this value is actually not difficult to obtain.  Let
$u_{+}$ and $u_{-}$ be two independent, scalar solutions of the
Sturm-Liouville problem (\ref{7}) (dropping the point interaction
term), vanishing at $+\infty$ and $-\infty$, respectively.  Therefore
we can write
\begin{equation}
g(0,0;E)={u_{+}(0)u_{-}(0) \over W[u_{+},u_{-}]},
\label{13}
\end{equation}
where $W[u_{+},u_{-}]$  is the Wronskian of the two solutions.  Defining
two real parameters
\begin{eqnarray}
K_L&=&{1\over \hbar v_z}\,\sqrt{ {1\over 4}
\,E_{GL}^2-(E-V_L)^2+v_\perp^2p_\perp^2 }, \nonumber \\
K_R&=&{1\over \hbar v_z}\,\sqrt{ {1\over 4}
\,E_{GR}^2-(E-V_R)^2+v_\perp^2p_\perp^2 },
\label{14}
\end{eqnarray}
the two independent solutions are $u_{+}=\exp (-K_R z)$ and $u_{-}=\exp
(K_L z)$ so that $g(0,0;E)=2/(K_R+K_L)$.  Using Eq.~(\ref{12}) one
finally obtains
\begin{equation} K_R+K_L={1\over\hbar
v_z}\sqrt{{1\over4}\,(E_{GR}-E_{GL})^2- (V_R-V_L)^2}.
\label{15}
\end{equation}
$K_R$ and $K_L$ should be real for obtaining exponentially decreasing
envelope-functions as $|z|\to \infty$ and then the gaps must overlap,
i.\ e.\ $(E_{GR}-E_{GL})^2/4>(V_R-V_L)^2$.  This solution agrees with
that previously proposed by Korenman and Drew \cite{Korenman}.  The
reader is referred to Ref.~\cite{Korenman} for a fully discussion of its
physical implications.  Here we stress the main advantages of using our
method.  First of all, we have restricted ourselves to the case of no
external potentials.  Nevertheless, it is clear that applied electric or
magnetic fields can be easily handled with minor modifications of the
equations.  Note that the crucial point is that one assumes that the
Klein-Gordon equation {\em without} the point interaction potential
arising from the abrupt interface can be solved exactly and the
corresponding Green functions is explicitly written out.  This is so for
a large variety of electric and magnetic field configuration, as pointed
out in Ref.~\cite{NC}.  Thus, for instance, it is possible to
investigate Landau levels in band-inverted heterojunctions in a rather
simple way, instead of using more elaborated mathematical treatments, as
those recently carried out by Aggasi \cite{Agassi2}.  In addition, it is
also possible to study confined Stark effect, a topic which remains open
in the literature.  The second aspect we remark is the fact that there
is no need to use an abrupt heterojunction model, simulated by a step
potential.  The only requirement is that $K_R^{-1}$ and $K_L^{-1}$ must
be much larger than the interface itself, an implicit assumption when
using the envelope-function formalism.  Qualitatively the profile of the
heterojunction is {\em soliton-like} \cite{Pan2} and, as a consequence,
it derivative is a sharply peaked function.  Thus the integral equation
(\ref{11}) can be solved by a limiting process, in analogous way to the
Dirac equation for sharply peaked functions approaching the
$\delta$-function limit \cite{JPA3}.  To conclude, we feel that the
approach we developed holds in a large variety of cases of practical
interest and it may help in a better understanding of interface states
in band-inverted heterojunctions.

\acknowledgments

This work is supported by UCM through project PR161/93-4811.

\end{document}